\documentclass[showpacs,amsmath,amssymb,pre]{revtex4}
\usepackage{graphics,color}
\begin{document}
\title{Large deviation techniques applied to systems with long-range interactions}
\author{Julien Barr{\'e}$^{1,2}$\thanks{E-mail: jbarre@cnls.lanl.gov},
  Freddy Bouchet$^{1}$\thanks{E-mail: Freddy.Bouchet@ens-lyon.fr}, Thierry
  Dauxois$^{1}$\thanks{E-mail: Thierry.Dauxois@ens-lyon.fr}, Stefano
  Ruffo$^{1,3}$\thanks{E-mail: ruffo@avanzi.de.unifi.it} }
\affiliation{1. Laboratoire de Physique, UMR-CNRS 5672, ENS Lyon,
46
  All\'{e}e d'Italie, 69364 Lyon c\'{e}dex 07, France\\
  2. Theoretical Division, Los Alamos National Laboratory, USA\\
 3. Dipartimento di Energetica, ``S. Stecco'' and CSDC, Universit{\`a} di
  Firenze, and INFN, via S. Marta, 3, 50139 Firenze, Italy }

\date{\today}

\begin{abstract}
  We discuss a method to solve models with {\it long-range interactions}
  in the microcanonical and canonical ensemble.  The method closely
  follows the one introduced by R.S. Ellis, Physica D 133, 106
  (1999), which uses large deviation techniques.  We show how it can
  be adapted to obtain the solution of a large class of simple models,
  which can show {\it ensemble inequivalence}.  The model Hamiltonian
  can have both {\em discrete} (Ising, Potts) and {\em
  continuous} (HMF, Free Electron Laser) state variables. This latter extension
  gives access to the comparison with dynamics and to the study of non-equilibrium 
  effects. We treat
  both infinite range and slowly decreasing interactions and, in
  particular, we present the solution of the $\alpha$-Ising model in
  one-dimension with $0\leq \alpha<1$. 
\end{abstract}

 \maketitle

A system with long-range interactions is characterized by an
interparticle potential $V(r)$ which decreases at large distances
$r$ slower than a power $r^{-\alpha}$ with $\alpha<d$, $d$ being
the dimension of the embedding space~\cite{Dauxois02LNP}.
Classical examples are self-gravitating~\cite{Padmanabhan90} and
Coulomb~\cite{Martin} systems, vortices in two-dimensional fluid
mechanics~\cite{Chavanis02houches},  wave-particles
interaction~\cite{Elskens02houches,Escande} and trapped charged
particles~\cite{gaspard}.  The behaviour of such systems is
interesting both from the dynamical point of view, because they
display peculiar quasi-stationary states that are related to the
underlying Vlasov equations~\cite{yoshi}, and from the static
point of view, because equilibrium statistical mechanics shows new
types of phase transitions and cases of ensemble
inequivalence~\cite{MukamelHouches}. In this paper we will
restrict ourselves to the second aspect.

In long-range interacting systems, essentially all the particles
contribute to the local field: the fluctuations around the mean
value are small because of the law of large numbers. This explains
qualitatively why the mean-field scaling, which amounts to let the
number of particles go to infinity at fixed
volume~\cite{limitperformed,spohn}, is usually extremely good.
However, let us remind that for long-range interacting systems,
microcanonical and canonical ensembles are not necessarily equivalent in the
mean-field limit~\cite{Thirring,Kiessling97,juBEG}.  Moreover,
because of the non additivity of the energy, the usual
construction of the canonical ensemble cannot be applied. This is
the reason why the microcanonical ensemble is considered by some
authors~\cite{Padmanabhan90,Gross01} as the only ``physically
motivated'' one.  Hence, it is extremely important to develop
rigorous techniques to solve non trivial physical models in the
microcanonical ensemble. One finds in books the solution for the
perfect gas, but generalizations to interacting particle systems
are difficult.

The goal of this paper is to advocate the use of large deviation
techniques as a tool to {\em explicitly} derive microcanonical and
canonical equilibrium solutions for a wide class of models.  As a
first step in this direction, we will discuss here a general
solution method to treat in full detail mean-field models without
short distance singularity.  Large deviation
techniques~\cite{Dembo,Ellis85} are nowadays widely used.  For
example, Michel and Robert~\cite{Michel94}, using these
techniques, rigorously derived the statistical mechanics of the
two-dimensional Euler equations. The method was later used by
Ellis \emph{et al.}~\cite{Ellis99} to study other two-dimensional
geophysical fluids.  The statistical mechanics of some models with
{\em discrete} variables has been recently obtained using large
deviation theory~\cite{barrethesis,MukamelHouches,Touchette2003}.
Besides presenting the solution method, we will show in this paper
how it can be applied to models whose Hamiltonian depends on {\em
continuous} state variables, like the so-called Hamiltonian
Mean-Field (HMF) model~\cite{Antoni95,Dauxois02}; the interest
being here to obtain solutions in the microcanonical ensemble of
models which display a Hamiltonian dynamics, opening the
possibility of studying also non-equilibrium effects.

We will first briefly introduce in Section~\ref{intro_grandes_dev}
the large deviation technique. In Section~\ref{Methode_generale}
we will recall the different steps of the mathematical framework
introduced in Ref.~\onlinecite{Ellis99}, necessary for a
systematic application to long-range interacting systems: we will
use the infinite range Potts model as a simple example.  In a
first instance, we will then treat infinite range models with
continuous variables. We will present in Section~\ref{exemple_HMF}
the solution of the HMF model~\cite{Antoni95} in the
microcanonical ensemble and, in the following
Section~\ref{Colson-Bonifaciomodel}, we will consider the
Colson-Bonifacio model of Free Electron Lasers
(FEL)~\cite{Bonifacio90}, as an example of relevance for physical
applications. In Section~\ref{Exemple_alphaIsing}, we will show
that these techniques can be applied also to cases where the
interaction is not infinite range but distance dependent; the
solution of the so-called $\alpha$-Ising model~\cite{dyson}, in
one dimension with $0\leq\alpha<1$, will be presented in full
detail, allowing us to discuss also the role of boundary conditions. Finally,
Section \ref{remarques} will be devoted to conclusions and
perspectives.

\section{Large deviation theory}
\label{intro_grandes_dev}

We will present in this section the main ideas behind large
deviation techniques, with an emphasis to applications. For a more
rigorous mathematical treatment, we direct the reader to
Refs.~\cite{Dembo,Ellis85}.

\subsection{The large deviation principle}
\label{What}

Let us consider the sample mean of $N$ independent real random
variables $X_k$ with the same distribution and zero average
\begin{equation} S_N=\frac{1}{N}\sum_{k=1}^{N}X_k\quad. \label{sum}
\end{equation}
The law of large numbers states that $S_N$ tends toward the
average $x=\langle X_k\rangle$,  namely $0$, when $N$
tends toward infinity. Moreover, if $X_k$ has a finite variance,
since all hypotheses of the central limit theorem  are fulfilled,
the probability distribution $P(\sqrt{N}S_N\in[x,x+dx])$ converges
towards a Gaussian, hence the fluctuations of $S_N$ are of order
$1/\sqrt{N}$. Typical questions of the large deviation theory are:
What is the behavior of the tails of the distribution? What is the
probability of a fluctuation of order one of $S_N$, {\it i.e.}
what is the value of $P(S_N\in[x,x+dx])$?

Let us be more specific by discussing the usual example of the
coin toss. We attribute to heads and tails of a coin the values
$X_k=+1$ and $X_k=-1$, respectively. The sum $S_N$ can take
($N+1)$ distinct $x$-values in the interval $[-1,1]$. For such
values, using simple combinatorial analysis, one easily derives
the probability distribution
\begin{equation}
P(S_N=x) = \frac{N!}{\displaystyle\left(\frac{(1+x)N}{2}\right)!\
\left(\frac{(1-x)N}{2}\right)!\ 2^N}\quad,
\end{equation}
which, using the Stirling's formula, can be approximated in the
large $N$ limit as
\begin{eqnarray}
\ln P(S_N=x) &\sim&
-N\left(\frac{(1+x)}{2}\ln{(1+x)}+\frac{(1-x)}{2}\ln{(1-x)}\right)
         \equiv -N I(x)\quad, \label{pgdpileouface}
\end{eqnarray}
which defines the function $I(x)$. More precisely, one can prove
that for any  interval $]x_1,\ x_2[ \subset[-1,1]$,
\begin{equation}
\lim_{N\to \infty}-\frac{1}{N}\ln P(x\in ]x_1,\ x_2[) =\max_{x\in
]x_1, \ x_2[} I(x)\quad.
\end{equation}
In the language of large deviation theory, one states that $S_N$
fulfills a {\em large deviation principle}, characterized by the
rate function~$I(x)$. If one interprets the coin toss experiment
as a microscopic realization of a chain of $N$ non-interacting
Ising spins, it is straightforward to prove that, in the
statistical mechanics vocabulary, $I(x)$ corresponds to the
negative of the Boltzmann entropy (divided by the Boltzmann
constant) of a state characterized by a fraction $x$ of up-spins.
This is a first simple example of the large deviation principle,
and the main purpose of this  paper is to present other examples
of its use for more complicated and physically relevant systems.

\subsection{Cram{\'e}r's theorem} \label{how}

Cram{\'e}r's theorem~\cite{Dembo} allows one to derive the probability
distribution $P(S_N\in[x,x+dx])$ in the large $N$-limit, providing
also a method to compute the rate function~$I(x)$. The theorem is
formulated for multi-dimensional and identically distributed random
variables $X_k$ $\in\mathbb{R}^d$, $d$ being the dimension of the
space of the variables. We will formulate the theorem in an
informal way, without emphasizing mathematical technicalities.

Let us define the function $\Psi({\lambda})$ as
\begin{equation}
\Psi({\lambda})=\langle e^{\lambda\cdot X}
\rangle\quad,\label{defpsi}
\end{equation}
where $\lambda\in\mathbb{R}^d$, ``$\cdot$'' is the usual scalar
product and $\langle \quad \rangle$ is the average over the common
probability distribution of the variables~$X_k$. Cram{\'e}r's
theorem states that, if $\Psi({\lambda})<\infty$, $\forall\lambda\in\mathbb{R}^d$,
 then the sample
mean $S_N$ satisfies the large deviation principle
\begin{equation}
\ln P(S_N\in[x,x+dx]) \sim -N I({x})\quad, \label{pgdcramer}
\end{equation}
with rate function $I({x})$ (${x}\in\mathbb{R}^d $) given by the
Legendre-Fenchel's transform of $\ln \Psi$,
\begin{equation}
I({x})=\sup_{{\lambda}\in\mathbb{R}^d}\left(\lambda\cdot
{x}-\ln{\Psi({\lambda})}\right)\quad. \label{Cramer}
\end{equation}
The formulation of the theorem
is not restricted to discrete random variables: this will be important for
our applications.

A heuristic proof of the theorem for the simplest case $X_k \in
\mathbb{R}$ goes as follows. The probability of obtaining $S_N=x$,
with $d\mu$ the common probability distribution of each
variable~$X_k$, is given by
\begin{eqnarray}
P(S_N\in[x,x+dx])  & = & \int \prod_{k=1}^Nd\mu(X_k) \
\delta(S_N-x)\quad.\label{heuristic}
\end{eqnarray}
This formula can also be interpreted as the volume of the
phase-space $(X_1,\cdots,X_N)$ under the microcanonical constraint
that $S_N=x$.  Using the Laplace transform of the Dirac's
$\delta$-function, one obtains
\begin{equation}
    P(S_N\in[x,x+dx])       =  \frac{1}{2\pi i}\int_{\Gamma} d\lambda \ e^{-N\lambda x}
      \int \prod_{k=1}^Nd\mu(X_k)\  e^{\lambda\sum_{k=1}^NX_k}\quad,
\end{equation}
where $\Gamma$ is a path on the complex $\lambda$-plane going from
$-i\infty$ to $+i\infty$, which crosses the real axis at a
positive value. Subsequent manipulations of this formula lead to
\begin{equation} P(S_N\in[x,x+dx])  =  \frac{1}{2\pi i} \int_{\Gamma} d\lambda\
e^{-N\lambda x} \left[\langle
  e^{\lambda X} \rangle \right]^N= \frac{1}{2\pi i} \int_{\Gamma} d\lambda\  e^{-N
  \left(\lambda x- \ln \langle e^{\lambda X} \rangle \right)}
  \stackrel{N\to \infty}{\simeq}e^{-NI(x)}\quad,
\end{equation}
where $I(x)$ is given in formula~(\ref{Cramer}) with $d=1$.  In
the last step, a large $N$ saddle-point approximation has been
performed.

Most of the results contained in this paper will be obtained using 
Cram{\'e}r's theorem, because the statistical variables of the models
we will consider are identically distributed  in space (mostly
on a lattice). In all cases the function $\ln \Psi$ is differentiable,
which also fulfils the hypotheses of the G{\"a}rtner-Ellis theorem~\cite{Dembo}.

\section{A general method}
\label{Methode_generale}

In this Section, we will describe the use of the large deviation
method to solve models with long-range interactions. As already
mentioned, Michel and Robert~\cite{Michel94} successfully used
large deviations techniques to rigorously prove the applicability
of statistical mechanics to two-dimensional fluid mechanics,
proposed earlier~\cite{Miller90,Robert91}. Ellis {\em et al.}~\cite{Ellis99} have
developed and generalized this approach to solve
two-dimensional geophysical systems with long-range interactions.
Here, we will adopt Ellis {\em et al.}'s approach, emphasizing the
different steps in the construction of thermodynamic functions.
The method will be exemplified discussing in detail the three-state
 Potts model with infinite range interactions.  This simple
example has been recently used as a toy model to illustrate
peculiar thermodynamic properties of long-range
systems~\cite{Ispolatov01a}. The diluted three-state Potts model
with short-range interactions has also been studied in connection
with ``small'' systems thermodynamics by Gross~\cite{Gross00}.

The Hamiltonian of the three-state Potts model is
\begin{equation}
 H_N = -\frac{J}{2N}\sum_{i,j=1}^{N} \delta_{S_i,S_j}~.
\label{Hpotts3}
\end{equation}
The $1/N$ prefactor is introduced in order to keep energy
extensive~\cite{Alternatively}. Each lattice site $i$ is occupied
by a spin variable $S_i$, which assumes three different states
$a$, $b$, or $c$. A pair of spins gives a ferromagnetic
contribution $-J$ ($J>0$) to the total energy if they are in the
same state, and no contribution otherwise. It is important to
stress that the energy sum is extended over {\em all} pairs
$(i,j)$: the interaction is infinite range.

The solution method consists of three steps.
\bigskip

{\bf Step 1~: Identifying  global variables}

Let $\Sigma_N$ be the phase-space of a $N$-particles system with
Hamiltonian
\begin{eqnarray}
H_N &:&\Sigma_N\to\mathbb{R}
\end{eqnarray}
and  $\omega_N \in\Sigma_N $ be a specific microscopic
configuration. The first step of the method consists in
associating to every microscopic configuration $\omega_N$, a
global (coarse-grained) variable, $\gamma(\omega_N)$. Then, a new
Hamiltonian $\widetilde{H}_N$ can be defined
\begin{eqnarray}
\label{approxH_N} H_N(\omega_N)=\widetilde{H}_N\left(\gamma
(\omega_N)\right)+R_N(\omega_N)\quad.
\end{eqnarray}
If one can neglect $R_N(\omega_N)$ with respect to
$\widetilde{H}_N$ in the large $N$-limit, then the Hamiltonian can
be expressed only in terms of the global variables. When
considering the above defined infinite range Potts model, the
appropriate global variable is
\begin{equation}\label{muvarPotts}
\gamma=(n_a,n_b,n_c)\quad,
\end{equation}
where $(n_a,n_b,n_c=1-n_a-n_b)$ are the fractions of spins in the
three different states $a,b,c$. In this case, the Hamiltonian
expressed in terms of the global variable is
\begin{equation}
\widetilde{H}_N =
-\frac{JN}{2}(n_a^2+n_b^2+n_c^2)\quad,\label{Hpottpoly}
\end{equation}
and coincides with the original Hamiltonian ${H}_N$ even at
finite~$N$. In Section~\ref{Exemple_alphaIsing}, we will discuss
the $\alpha$-Ising model, for which~$R_N$ does not vanish.

The global variable $\gamma $ is of finite dimension in our
example, but could be of infinite dimension in other cases. For
instance, $\gamma $ could correspond to a local mass density in a
gravitational system, or a coarse-grained vorticity density in 2D
turbulence.

Although this type of redefinition of the Hamiltonian might appear
to be possible in all cases, this is not true. For instance, in
the case of short-range interactions, even after defining a local
density of a physical quantity (e.g. magnetization), when
performing the large $N$-limit, no general argument exists to
neglect~$R_N$. As a consequence, in such a case, the local density (e.g. the local
magnetization) is not the appropriate macroscopic variable. 
However we will argue that such a procedure is,
instead, viable in general for systems with long-range
interactions on a lattice. Keeping the lattice spacing finite is
important in order to regularize possible short distance
singularities.

\bigskip {\bf Step 2~: Deriving an entropy functional for the global
  variables.}

Because the mean-field variables are not equiprobable, the number of
microscopic configurations leading to a given value of $\gamma $
does depend on $\gamma $ itself. Then, one can define an entropy
functional~$s(\gamma )$,
\begin{equation}
s(\gamma )=\lim_{N\to\infty }\frac{1}{N} \ln  \Omega_N(\gamma
)=-I(\gamma)+\ln {\cal N} \quad,\label{largedev}
\end{equation}
where the Boltzmann constant has been set to unity,
$\Omega_N(\gamma )$ is the number of microscopic configurations
corresponding to a given value of~$\gamma $, $I(\gamma)$ is the
rate function and $ {\cal N}=\int d\gamma
\Omega_N(\gamma )$ is the total number of states. This is where
large deviation theory applies; not only in proving that such an
entropy functional exists, but also in providing a procedure to
derive it explicitly.

If, besides the dynamical variables which contribute to the global
ones, the Hamiltonian depends also on a number of variables $n_v$
which is small with respect to $N$, it is easy to prove that these
additional variables  contribute to the entropy for a negligible
term, proportional to $n_v/N$.

For the infinite range Potts model, it is possible to derive the
entropy functional $s(\gamma)$ directly, using combinatorial
arguments and the Stirling approximation in the large $N$-limit.
However, let us follow instead the procedure given by Cram{\'e}r's
theorem as explained in Section~\ref{how}.
Expression~(\ref{muvarPotts}) for $\gamma $, which is identified
with $x$ in Section~\ref{how}, can be rewritten as
\begin{equation}
\gamma =\left(\frac{1}{N}\sum_i \delta_{S_i,a},\frac{1}{N}\sum_i
\delta_{S_i,b},\frac{1}{N}\sum_i \delta_{S_i,c} \right)\quad.
\end{equation}
The local random variables with common probability distribution
are here
\begin{equation} X_k=\left(\delta_{S_k,a}, \delta_{S_k,b}, \delta_{S_k,c}\right)\quad.
\end{equation}
Hence, the  generating function $\Psi$ is given by
\begin{eqnarray}
\Psi(\lambda_a,\lambda_b,\lambda_c) &=&
\frac{1}{3}\sum_{S=a,b,c}\left(e^{\lambda_a
\delta_{S,a}+ \lambda_b \delta_{S,b}+ \lambda_c \delta_{S,c}}\right) \\
&=&  \frac{1}{3}
\left(e^{\lambda_a}+e^{\lambda_b}+e^{\lambda_c}\right)\quad.
\end{eqnarray}
The large deviation functional is
\begin{equation}
I(\gamma )=\sup_{\lambda_a,\lambda_b,\lambda_c}\left(\lambda_a
n_a\:+ \lambda_b n_b\:+ \lambda_c
n_c\:-\ln{\Psi(\lambda_a,\lambda_b,\lambda_c})\right)\quad.
\end{equation}
This variational problem can be solved exactly, giving
$\lambda_\ell =\ln n_\ell$ with $\ell=a,b,c$. Hence,
\begin{eqnarray}
I(\gamma ) &=&n_a\ln n_a+n_b\ln n_b+(1-n_a-n_b)\ln(1-n_a-n_b)+\ln
3\quad.
\end{eqnarray}
Thermodynamic entropy density is given by $s(\gamma )=-I(\gamma
)+\ln {\cal N}$, where the normalization factor is ${\cal N}=3$
for the Potts model example, which recovers the result of the
combinatorial approach.

  \bigskip

{\bf Step 3~: Microcanonical and canonical variational problems}
\bigskip

To obtain the entropy as a function of energy density
$\varepsilon$, {\it i.e.} to solve a model in the microcanonical
ensemble, after performing steps 1 and 2 of the solution method, 
one has to find the solution of the following variational
problem~\cite{Ellis99}
\begin{equation}
\label{varprobleme_micro} S(\varepsilon)=\sup_{\gamma } \left(
s(\gamma  )\: | \: H(\gamma )=\varepsilon\right)\quad,
\end{equation}
where
\begin{equation}
H(\gamma)=\lim_{N\to\infty} \frac{\widetilde{H}_N(\gamma
)}{N}\quad.
\end{equation}
This result is exact but it corresponds also to an ``intuitive''
mean-field solution.  Finally, let us notice that the total
entropy density is intensive: there is no difference in this
respect with short-range interacting systems.  When other
conserved quantities exist, they must be taken into account when
solving the variational problem in formula~(\ref{varprobleme_micro}).
Specific examples will be discussed in sections~(\ref{exemple_HMF})
and~(\ref{Colson-Bonifaciomodel}).

\begin{figure}
\resizebox{0.5\textwidth}{!}{\includegraphics{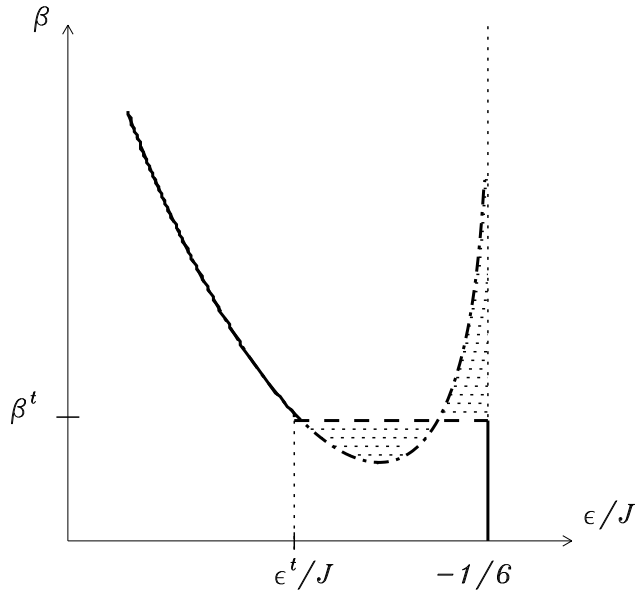}}
\caption{Caloric curve (inverse temperature vs. energy density)
 of the three states infinite range Potts
  model.  The canonical solution is represented by a solid line.
The microcanonical solution coincides with the canonical one for
$\varepsilon\leq \varepsilon^t$ and is instead indicated by the
dash-dotted line for $\varepsilon^t\leq \varepsilon<-J/6$. The
increasing part of the microcanonical dash-dotted line corresponds
to a negative specific heat region. In the canonical ensemble, the
model displays a first order phase transition at $\beta^t$. The
two dotted regions bounded by the dashed line and by the
microcanonical dash-dotted line have the same area (Maxwell's
construction).} \label{potts_fig}
\end{figure}

For the infinite range Potts model~(\ref{Hpotts3}), the
variational problem  is
\begin{equation}
\label{entropie_potts} S(\varepsilon) =\sup_{n_a,n_b} \Bigl(
-n_a\ln n_a-n_b\ln n_b-(1-n_a-n_b) \ln(1-n_a-n_b)\ \Bigl|
       -\frac{J}{2}\left(n_a^2+n_b^2+(1-n_a-n_b)^2\right)=\varepsilon \Bigr)\quad.
\end{equation}
This variational problem can be solved numerically.  The
microcanonical inverse temperature
${\beta}(\varepsilon)=dS/d\varepsilon$ can then be derived: it is
shown in Fig.~\ref{potts_fig} in the allowed energy range
$[-J/2,-J/6]$. Ispolatov and Cohen~\cite{Ispolatov01a} have
obtained the same result by determining the
density of states. A negative specific heat region appears in the
energy range $[-0.215\, J,-J/6]$.

Let us now consider  the canonical ensemble. The partition
function, written in terms of the common probability distribution
of the phase-space variables, is
\begin{eqnarray}
Z({\beta},N)  &=& \int \prod_{k=1}^N d \mu (X_k)\,
e^{\displaystyle -{\beta} {H}_N}
\end{eqnarray}
where $\mu$ is the probability density of $X_k$. This is not the
usual partition function, but differs from it only for a constant factor,
which counts the number of states. Let us remark that we have also
used the letter $\beta$ for the microcanonical inverse
temperature. We will comment specifically when  the microcanonical
inverse temperature differs from the canonical one due to ensemble
inequivalence.

Approximating  Hamiltonian $H_N$ in the large $N$-limit with the
one expressed in terms of the global variable $\gamma$, one gets
\begin{eqnarray}
Z(\beta,N) &\stackrel{N\to \infty}{\simeq} & \frac
{\displaystyle\int d\gamma\, \Omega_N(\gamma)\,e^{\displaystyle
-\beta H(\gamma )}} {\displaystyle\int d\gamma\, \Omega_N(\gamma)}
\quad.\label{equation32}
\end{eqnarray}

For infinite range models, formula (\ref{equation32}) is  exact
for all $N$ because the rest $R_N$ vanishes. Using formula
(\ref{largedev}), one obtains
\begin{eqnarray}
Z({\beta},N) &\simeq &  \int  d\gamma \ e^{\displaystyle
-N\left(-s(\gamma)+{\beta}{{H}(\gamma)}{}\right)}\quad.
\end{eqnarray}
Applying the saddle point method, the partition function is
rewritten as
\begin{eqnarray}
Z(\beta,N) &\simeq & e^{\displaystyle -N F(\beta)}\quad,
\end{eqnarray}
where the ``free energy'' $F(\beta)$ is obtained solving the
variational problem
\begin{equation}
\label{varprobleme_cano} F(\beta)=\inf_{\gamma } \left( \beta
H(\gamma )-s(\gamma )\right)\quad.
\end{equation}
Our ``free energy'' is the usual free energy multiplied by the
inverse temperature. This helps because physical states will
correspond to minima of such free energy also for negative
inverse temperatures. 

In the case of the infinite range three-state Potts model, the
canonical  free energy can be explicitly derived solving the
following variational problem
\begin{equation}
\label{elibre_potts} F(\beta)=\inf_{n_a,n_b,n_c} \Bigl( n_a\ln
n_a+n_b\ln n_b+n_c \ln n_c -\frac{\beta
J}{2}\left(n_a^2+n_b^2+n_c^2\right)\left|n_a+n_b+n_c=1\right.\Bigr)\quad.
\end{equation}
To obtain the caloric curve, one has to compute $\varepsilon={d F/
d \beta }$. Fig.~\ref{potts_fig} shows that at the  canonical
transition inverse temperature $\beta^t\simeq2.75$, corresponding to
the energy $\varepsilon^t/J\simeq-0.255$, a first order phase
transition appears, with an associated latent heat. The low energy
``magnetized'' phase becomes unstable, while the high energy
``homogeneous'' phase, which has the constant energy density,
$\varepsilon/J=-1/6$, is stabilized. In Fig. \ref{potts_fig},
the two dotted regions have the same area, respecting Maxwell's
construction.  At the inverse transition temperature, there is
also a jump in the global variables $(n_a,n_b,n_c)$ which are the
order parameters of the model.

This extremely simple example shows already {\em ensemble
inequivalence}. In the microcanonical ensemble, there is no phase
transition and the specific heat becomes negative. On the other
hand, in the canonical ensemble, there is a first order phase
transition with a latent heat. The caloric curves do not coincide.
We observe that in the energy range of ensemble inequivalence,
microcanonical temperatures,
$(dS/d\varepsilon)^{-1}$, do not coincide with any canonical
one.

This is an example of the more general fact that entropy
$s(\varepsilon)$ is  not always the Legendre-Fenchel transform of
the free energy $F(\beta)$. A general discussion of ensemble
inequivalence, both at the thermodynamic level and at the level
of equilibrium macrostates, is provided in
Refs.~\cite{Ellis99,Touchette2003}, while a classification of
phase transition and of ensemble inequivalence situations is
reported in Ref.~\cite{BouchetBarre}.

This concludes the general presentation of the different steps of
the method to derive the statistical mechanics of long-range
interacting systems.

\section{Examples}

In this Section, we will discuss the application of the large
deviation method to two examples which share the difficulty of
computing the entropy for a phase-space with {\em continuous}
variables. When presenting the method, we have discussed in
parallel its application to a model with {\em discrete} variables,
the infinite range three-state Potts model, which however could
have been solved by direct states counting~\cite{Ispolatov01a}.
This latter approach cannot be used when the variables are
continuous. Obtaining microcanonical entropy often implies the
solution of too complicated integrals, and indeed one does not
find many examples of such solutions in the literature. On the
contrary, the large deviation method is not restricted to discrete
variables and we will show that it can even be simpler to use in
such a case.

\subsection{The Hamiltonian Mean Field model} \label{exemple_HMF}

The Hamiltonian Mean Field  (HMF) model~\cite{Dauxois02,Antoni95}
is defined by the following Hamiltonian
\begin{equation}
H_N=\sum_{i=1}^{N}\frac{p_i^2}{2}+\frac{C}{2N}
\sum_{i,j}\cos(\theta_i-\theta_j)\quad, \label{Ham_HMF}
\end{equation}
where $\theta_i\in[0,2\pi[$ is the position (angle) of the $i$-th
article on a circle and $p_i$ the corresponding conjugate
variable. This system can be seen as representing particles moving
on a unit circle interacting via an infinite range attractive
($C<0$) or repulsive ($C>0$) cosine potential or, alternatively,
as classical XY-rotors with infinite range ferromagnetic ($C<0$)
or antiferromagnetic ($C>0$) couplings. The renormalization factor
$N$ of the potential energy is kept not only  for historical
reasons, but also because it simplifies the derivation of the
variational problems and makes the problem well defined. As we
will see, this implies that the usual energy
per particle and temperature are well defined in the $N\to\infty$ 
limit. 
In the literature, some authors have treated the case in which the
energy is not extensive. This leads to different thermodynamic
limit behaviors~\cite{toral,celia}.

The canonical solution of this model has been derived using the
Hubbard-Stratonovich transformation~\cite{Antoni95}.  The
microcanonical solution has been heuristically obtained, under the
hypothesis of concave entropy in Ref.~\cite{antoniHinrichsenruffo}
and in a different form in Ref.~\cite{velasquez}.  In this
section, we will derive both solutions with no additional
hypothesis. We will verify that the two ensembles give equivalent
predictions.

\medskip
\textbf{Step 1:} Hamiltonian~(\ref{Ham_HMF}) can be rewritten as
\begin{equation}
H_N=\sum_{i=1}^{N}\frac{p_i^2}{2}+\frac{NC}{2}(M_x^2+M_y^2)\quad,
\label{Ham_HMF2}
\end{equation}
where the magnetization is
\begin{equation} \mathbf{M}=M_x+i M_y=\frac{1}{N}\:\sum_k
e^{i\theta_k}\quad,\quad\mbox{with}\quad M=|\mathbf{M}|\quad.
\label{definitiondeM}
\end{equation}
 By a direct inspection of
Hamiltonian~(\ref{Ham_HMF2}), one can identify the global
quantities $u=\frac{1}{N}\sum_i p_i^2$, $M_x$ and $M_y$. Moreover,
since $v=\frac{1}{N}\sum_i p_i$ is a conserved quantity with
respect to the dynamics defined by Hamiltonian~(\ref{Ham_HMF}), it
will be included in the global variable. Hence,
\begin{equation}
\gamma =(u,v,M_x,M_y)\quad.
\end{equation}
The Hamiltonian in terms of the global variable is
\begin{equation}
H(\gamma)=\frac{1}{2}(u+CM^2)\quad.\label{defdeh}
\end{equation}

\medskip
 \textbf{Step 2:}
The vector of local variables is $X_k=(p_k^2,p_k,\cos
\theta_k,\sin \theta_k)$. The generating function is
\begin{eqnarray} \Psi(\lambda_u,\lambda_v,\lambda_c,\lambda_s)&=&
\left \langle e^{\displaystyle \lambda_up^2+\lambda_vp+
\lambda_c\cos \theta+\lambda_s\sin\theta } \right \rangle \\
&\sim&e^{-\lambda_v^2/4\lambda_u}\sqrt{\frac{\pi}{-\lambda_u}}\,
I_0\left(\sqrt{\lambda_c^2+\lambda_s^2}\right)
\quad,\label{formulepourpsi}
\end{eqnarray}
where $I_0$ is the modified Bessel function of order~0. In the
last expression, we have not reported the constant factor
$\int_\Sigma dp\,d\theta$ which is  finite because the domain of
integration $\Sigma$ is bounded due to the finiteness of the
energy. The sign~$\sim$ indicates that formula~(\ref{formulepourpsi}) 
is valid only at leading order; indeed, as
$\Sigma\neq\mathbb{R}\times[0,2\pi[$, 
neither the Gaussian nor the Bessel functions are fully exact.

The large deviation functional is then given, apart from trivial
constants, by
\begin{eqnarray}
\label{equi}
I(\gamma)=\sup_{\lambda_u,\lambda_v,\lambda_c,\lambda_s}
\left[\lambda_uu+\lambda_vv+\lambda_cM_x+\lambda_sM_y-\ln
I_0\left(\sqrt{\lambda_c^2+\lambda_s^2}\right)+\frac{\lambda_v^2}{4\lambda_u}+\frac{1}{2}\ln
(-\lambda_u) \right]\quad.
\end{eqnarray}
This variational problem can be solved for the ``kinetic''
subspace $(\lambda_u,\lambda_v)$ separately from the
``configurational'' one $(\lambda_c,\lambda_s)$. The entropy as a
function of the global variable is then
\begin{eqnarray}
 s(\gamma)=s_{kin}(u,v)+s_{conf}(M)\quad,
\end{eqnarray}
where
\begin{eqnarray}
s_{kin}(u,v)&=&\frac{1}{2}\ln{(u-v^2)}+\mbox{const}\label{scin}\\
s_{conf}(M)&=&-\sup_{\lambda}\left[\lambda M-\ln
I_0(\lambda)\right]=-\overline{\lambda} M+ \ln
I_0(\overline{\lambda} )\quad,\label{sconf}
\end{eqnarray}
with $\lambda= \sqrt{\lambda_c^2+\lambda_s^2}$ and
$\overline{\lambda}$ the solution of the variational problem in
(\ref{sconf}). Let us remark that Cauchy-Schwarz inequality
implies that $u\geq v^2$.

\medskip
\textbf{Step 3:} It is therefore possible to derive the
microcanonical variational problem
\begin{equation}\label{casmicrocanonique}
S(\varepsilon,v)=\sup_{M,u } \left[\frac{1}{2}\ln (u-v^2)
+s_{conf}(M)\: \Biggr| \frac{1}{2}u+\frac{C}{2}M^2=\varepsilon,
v=\mbox{const} \right]\quad,
\end{equation}
and the canonical one
\begin{eqnarray}\label{cascanonique}
F({\beta})&=&\inf_{M,u,v }
\left[\frac{{\beta}}{2}\left(u+CM^2\right) - \frac{1}{2}\ln
(u-v^2) -s_{conf}(M) \right]\quad.
\end{eqnarray}

We show here the solution of both  variational problems in the
case $v=0$ and $C=-1$, for which a second order phase transition
appears. Shifting $v$ to non-vanishing values does not produce
anything new, whereas the second order phase transition disappears
for positive values of $C$.

Solving the $\sup$ condition in Eq.~(\ref{sconf}) leads to the
consistency equation ${M}=I_1(\lambda)/I_0(\lambda)\equiv m(\lambda)$,
which determines the optimal value of $\lambda$,
$\overline{\lambda}=m^{-1}({M})$. Using the energy constraint,
Eq.~(\ref{casmicrocanonique}) can be rewritten as
\begin{equation}\label{casmicrocanoniquerewritten}
S(\varepsilon)=\sup_{M\in[0,1[} \left[s(M,\varepsilon)=
\frac{1}{2} \ln (2\varepsilon+ M^2) +s_{conf}(M) \right]\quad.
\end{equation}
In order to determine the global maximum of $S$, let us note that
in Eq.~(\ref{casmicrocanoniquerewritten}) $M$ takes values in the
interval $[0,1[$. Moreover, an asymptotic expansion shows that
$s(M,\varepsilon)\stackrel{M\to1}{\sim} (1/2)\ln (1-M)$,
which diverges to $-\infty$ when $M$ tends to 1. Thus, a global
maximum of the continuous function~$s$ with respect to $M$ exists
and is attained inside the interval $[0,1[$. To determine this
maximum one has first to solve the extremal condition
\begin{equation}
\frac{\overline{M}}{2\varepsilon+\overline{M}^2}=-s_{conf}'(\overline{M})=
m^{-1}\left(\overline{M}\right)\quad. \label{conditionaciter}
\end{equation}
The unique solution of this equation is $\overline{M}=0$ for
$\varepsilon>1/4$, while a non vanishing magnetization solution
bifurcates from it at $\varepsilon=1/4$, originating the second
order phase transition. One can indeed show that
\begin{equation}\label{form47}
s(M,\varepsilon)=\frac{\ln(2\varepsilon)}{2}+\left(\frac{1}{4\varepsilon}-1\right)M^2+o\left(M^4\right)\quad,
\end{equation}
which clarifies the stability change at $\varepsilon=1/4$ of the
$\overline{M}=0$ solution. Formula~(\ref{form47}) was also
obtained  in Ref.~\cite{BouchetPRE} in connection with the
fluctuations of the magnetization.

\begin{figure}
\resizebox{0.95\textwidth}{!}{\includegraphics{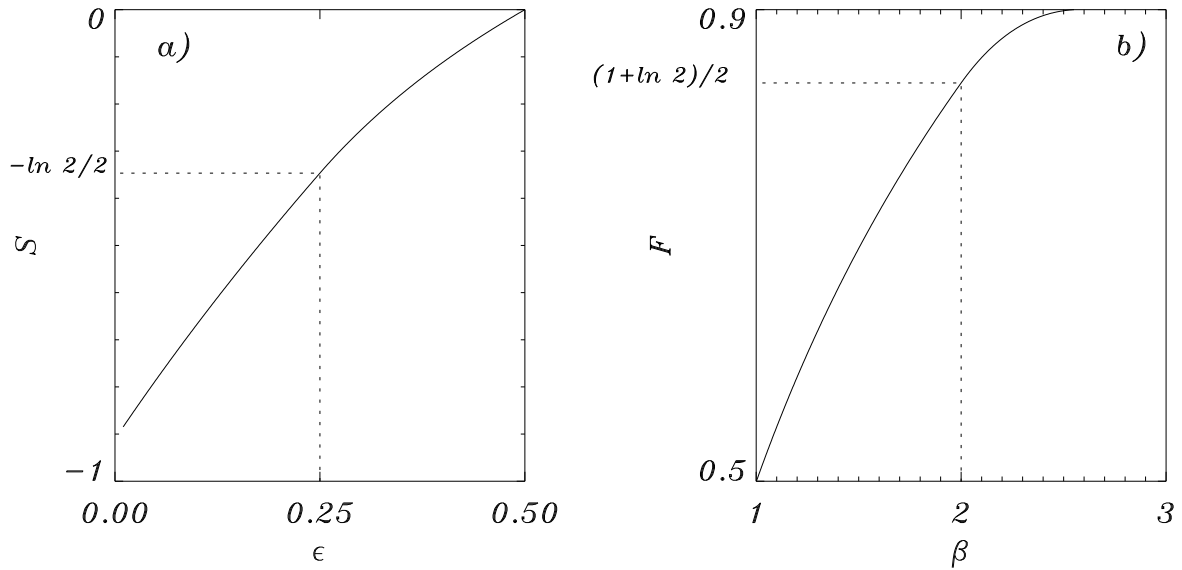}}
\caption{Entropy versus energy (a) and free energy versus inverse
  temperature (b) for the HMF model~(\ref{Ham_HMF2}) with $C=-1$ and
  $v=0$. The dotted lines are traced at the phase transition point.}
\label{hmfentropie}
\end{figure}

\begin{figure}
\resizebox{0.95\textwidth}{!}{\includegraphics{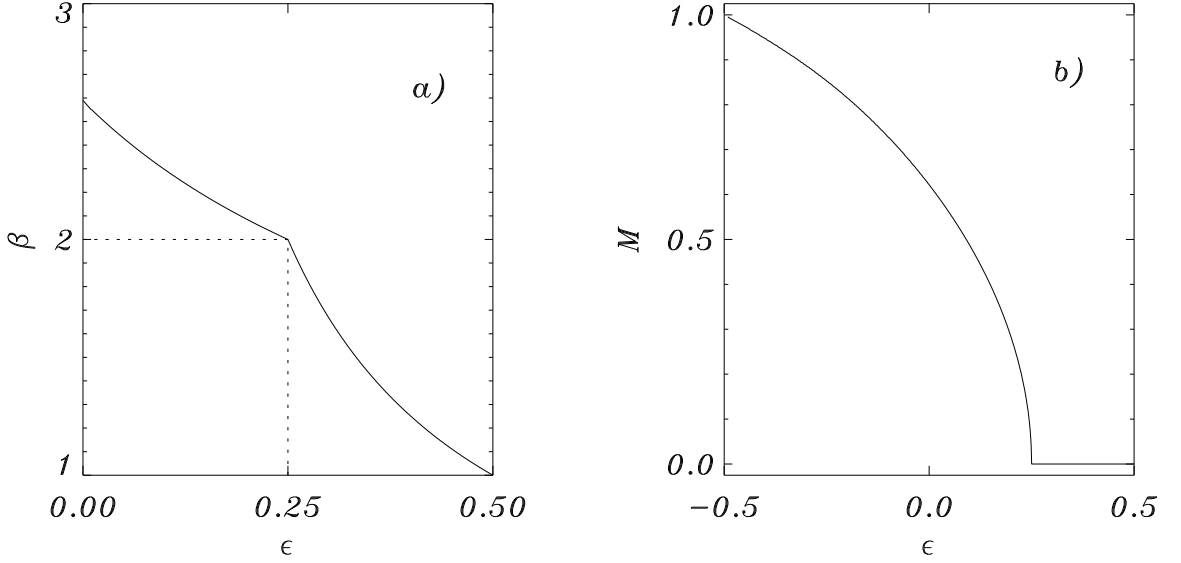}}
\caption{Inverse temperature versus energy (a) and magnetization
versus energy (b) for the HMF model~(\ref{Ham_HMF2}) with $C=-1$
and $v=0$. The dotted lines are traced at the phase transition
point.} \label{beteeth}
\end{figure}

Let us remark that $(d S)/(d\varepsilon)$ is nothing but the
inverse of twice the kinetic energy, which is the usual
microcanonical temperature. Moreover,
condition~(\ref{conditionaciter}) coincides, as expected, with the
consistency relation derived in the canonical
ensemble~\cite{Antoni95}.  In Fig.~\ref{hmfentropie}, we show the
full dependence of the entropy on energy and the graph of the free
energy versus the inverse temperature, obtained by solving numerically
the consistency equation~(\ref{conditionaciter}) in the low energy
(temperature) range.  In Fig.~\ref{beteeth}, we plot the caloric
curve and the dependence of the order parameter on energy: the two
ensembles give the same predictions because the entropy is
concave.

In Appendix A, we discuss a different way of obtaining the
microcanonical solution, which treats the kinetic part of the
energy in a more traditional way, like, for instance, in
gravitational dynamics~\cite{binneytremaine}.  The method showed
in this section is of more general applicability.

\subsection{The Colson-Bonifacio model for the Free Electron Laser}
\label{Colson-Bonifaciomodel}

In the linear Free Electron Laser (FEL), a relativistic electron
beam propagates through a spatially  periodic magnetic field,
interacting with the co-propagating electromagnetic wave; lasing
occurs when the electrons bunch in a subluminar beat
wave~\cite{Bonifacio90}. Scaling away the time dependence of the
phenomenon and introducing appropriate variables, it is possible
to catch the essence of the asymptotic state by studying the
classical Hamiltonian
\begin{equation}
H_N=\sum_{j=1}^N\frac{p_j^2}{2} -N \delta A^2 +2 A \sum_{j=1}^N
\sin(\theta_j-\varphi) \label{eq:Hamiltonien}\quad.
\end{equation}
The $p_i$'s represent the velocities relative to the center of
mass of the $N$ electrons and the conjugated variables~$\theta_i$
characterize their positions with respect to the co-propagating
wave.  The complex electromagnetic field variable, $\mathbf{A}=A\,
e^{i\varphi}$, defines the amplitude and the phase of the
dominating mode ($\mathbf{A}$ and $\mathbf{A}^\star$ are conjugate
variables). $\delta$ is a parameter which measures the average
deviation from the resonance condition. In addition to the
``energy'' $H$, the total momentum $P=\sum_j p_j + NA^2$ is also a
conserved quantity. Most of the studies of this model have
concentrated on the numerical solution of Hamiltonian
(\ref{eq:Hamiltonien}), starting from initial states with a small
field $A$ and the electrons uniformly distributed with a small
kinetic energy.  Then, the growth of the field has been observed
and its asymptotic value determined from the numerics. Our study
below allows to find the asymptotic value of the field
analytically.

\medskip \textbf{Step 1:} Similarly to the HMF case,
Hamiltonian~(\ref{eq:Hamiltonien}) can be rewritten as
\begin{eqnarray}
H_N\simeq  NH(\gamma) &=& N\left(\frac{u}{2}-\delta A^2 +2A
\left(-M_x\sin{\varphi} +M_y\cos{\varphi} \right) \right)
\end{eqnarray}
where $M_x$, $M_y$, $u$ and $v$ have been defined in
Eq.~(\ref{definitiondeM}) and following. Defining the phase of the
mean field~$\varphi^{\prime}$ as
$M_x+iM_y=M\exp{(i\varphi^{\prime})}$, the global variable is
$\gamma =(u,v,M,\varphi',A, \varphi)$.

\medskip \textbf{Step 2:} As remarked in Step 2 of
Section~\ref{Methode_generale},
 the contribution to the
entropy of the two field variables $A$, $ \varphi$, is negligible
(of order $1/N$). Hence, the $\Psi $ function reduces to the one
of the HMF model, see formula~(\ref{formulepourpsi}). Finally, one
obtains the same contributions to the kinetic and configurational
entropies, as shown in formulas~(\ref{scin}) and~(\ref{sconf}).

\medskip \textbf{Step 3:}  Defining the total momentum density as $\sigma=P/N$,
the microcanonical variational problem to be solved is
\begin{equation}
\label{eq:entropieLEL} S(\varepsilon,\sigma,\delta )=
 \sup_{\gamma }\Biggl[
\frac{1}{2}\ln{(u-v^2)} +s_{conf}(M) \, \Biggr| \, \varepsilon=
\frac{u}{2}+2A M\sin{\left(\varphi^{\prime}-\varphi\right)}
-\delta A^2, \, \sigma=v+A^2 \Biggr]\quad.
\end{equation}

Using the constraints of the variational problem, one can express
$u$ and $v$ as functions of the other variables, obtaining the
following form of the entropy
\begin{equation}
\label{eq:entropieLELbb} S(\varepsilon,\sigma,\delta )=  \sup_{A,
\varphi,M,  \varphi'  }\Biggl[
\frac{1}{2}\ln{\left[2\left(\varepsilon-\frac{\sigma^2}{2}\right)-4AM\sin\left(\varphi^{\prime}-\varphi\right)
+2(\delta-\sigma)
    A^2-A^4\right]} +s_{conf}(M)
 \Biggr]\quad.
\end{equation}
The extremization over the variables $\varphi$ and
$\varphi^{\prime}$ is straightforward, since by direct inspection
of formula (\ref{eq:entropieLELbb}), it is clear that the entropy
is maximized when $\varphi^{\prime}-\varphi=-\pi/2$. Then
\begin{equation}
\label{eq:entropieLEL2} S(\varepsilon,\sigma,\delta)=\sup_{A,  M
}\Biggl[
\frac{1}{2}\ln{\left[2\left(\varepsilon-\frac{\sigma^2}{2}\right)+4AM
+2(\delta-\sigma)
    A^2-A^4\right]} +s_{conf}(M) \Biggr]\equiv  \sup_{A, M} s(A,M)\quad.
\end{equation}
The non zero $\sigma$ case can be reduced to the vanishing
$\sigma$ problem using the identity
$S(\varepsilon,\sigma,\delta)=S(\varepsilon-\sigma^2/2,0,\delta-\sigma)$.
From now on we will discuss only the zero momentum case, changing
$\varepsilon+\sigma^2/2\to \varepsilon$ and
$\delta+\sigma\to\delta$. This has also a practical interest,
because it is the experimentally relevant initial condition
\cite{FELPRE}.

The conditions for having a local stationary point are
\begin{eqnarray}
\frac{\partial s}{\partial A} &=& \frac{2\left(\delta A-A^3
+M\right)}
{2\varepsilon+2\delta A^2+4AM-A^4}=0\quad, \label{eq:derivee1a} \\
\frac{\partial s}{\partial M} &=& \frac{2A} {2\varepsilon+2\delta
A^2+4AM-A^4} -m^{-1}(M)=0\quad, \label{eq:derivee1b}
\end{eqnarray}
where $m^{-1}$ is defined in formula (\ref{conditionaciter}).  It
is clear that $M=A=0$ is a solution of  conditions
(\ref{eq:derivee1a}) and (\ref{eq:derivee1b}): it exists only for
positive $\varepsilon$. We will limit ourselves to study its
stability. It must be remarked that this is the typical initial
condition studied experimentally in the FEL: it corresponds to
having a beat wave with zero amplitude and the electrons
uniformly distributed. The lasing phenomenon is revealed by an
exponential growth of both $A$ and the electron bunching 
parameter $M$.

The second order derivatives of the entropy $s(A,M)$, computed on
this solution, are
\begin{eqnarray}
\frac{\partial^2 s}{\partial A^2}(0,0) &=& \frac{\delta}{e}, \quad
\frac{\partial^2 s}{\partial m^2}(0,0) = -2, \quad
\frac{\partial^2 s}{\partial A \partial m}(0,0) =
\frac{1}{\varepsilon}\quad.
\end{eqnarray}
The two eigenvalues of the Hessian are the solutions of the
equation
\begin{equation}
x^2-x\left(-2+\frac{\delta}{\varepsilon}\right)-\frac{2\delta}{\varepsilon}-\frac{1}{\varepsilon^2}=0\quad.
\end{equation}
The stationary point is a maximum if the roots of this equation
are both negative. This implies that their sum ${\cal
S}=(-2+{\delta}/{\varepsilon})$ is negative and their product
${\cal P}=-{2\delta}/{\varepsilon}-{1}/{\varepsilon^2}$ is
positive.
 Recalling that we
restrict to positive $\varepsilon$ values, the condition for the sum to
be negative is $\varepsilon>\delta/2$ and the one for the product
to be positive is $\varepsilon>-1/(2\delta)$ with $\delta<0$. The
second condition is more restrictive, hence the only region where
the solution $M=A=0$ exists and is stable is
$\varepsilon>-1/(2\delta)$ with $\delta<0$. When crossing the line
$\varepsilon=-1/(2\delta)$ ($\delta<0$), a non zero bunching
solution ($M\neq0$) originates continuously from the zero bunching
one, producing a second order phase transition. This analysis
fully coincides with the one performed in the canonical ensemble
in Ref. \cite{Firpo00}.

The maximum entropy solution in the region complementary to the
one where the zero bunching solution is stable can be
obtained~\cite{FELPRE} by solving numerically
Eqs.~(\ref{eq:derivee1a}) and~(\ref{eq:derivee1b}). This
corresponds to having a non zero field intensity and bunching.

We have not completed in this case the study of the global
stability of the different solutions, but we think that, in view
of the possibility to map this model exactly onto the HMF model
(see Appendix B), no surprise is expected and that the study
presented here should fully represent all physical solutions.

\section{The Ising model with $1/r^{\alpha}$ interactions}
\label{Exemple_alphaIsing}

All the models that have been considered above are infinite range:
interactions are independent of the distance and the energy can
therefore be written {\em exactly} in terms of global variables.
This is no more valid for several important and physically
relevant cases. Let us mention in particular the $1/r$ interaction
law for gravity and Coulomb systems and the logarithmic
interaction for two-dimensional turbulence. A generalized
$1/r^{\alpha}$ interaction has been also introduced and the
resulting phase transitions have been
analyzed~\cite{ISPOCohenPRL}.  In all these cases the interaction
is singular at short-range.  Interesting toy models (which
generalize the HMF model) without a short distance singularity
have been recently
proposed~\cite{Anteneodo99,Anteneodo97,Campa00,Vollmayr01}.  In
these latter models, XY spins are put on a lattice and interact
through a slowly decreasing non integrable $1/r^{\alpha}$ law
($\alpha<d$).  For what concerns studies in the microcanonical
ensemble, is has been shown~\cite{Anteneodo99} that the
thermodynamic behavior of these models is independent of the
$\alpha$ exponent, after one has adopted an appropriate
renormalization of energy and temperature.  Salazar \emph{et
al.}~\cite{Salazar02} have studied numerically the problem using
microcanonical Monte-Carlo simulations. Besides confirming the
scaling properties with $\alpha$, they have also shown that the
number of states is of order $\exp(N)$.  The exact solution of
these models in the canonical ensemble has been obtained by Campa
\emph{et al.}~\cite{Campa00} and Vollmayr-Lee and Luijten
\cite{Vollmayr01}. The scaling of the magnetization and of the
energy curve with the $\alpha$-exponent has been exhibited in full
detail.

In this section, we will present a microcanonical solution of the
one dimensional $\alpha$-Ising model, whose Hamiltonian is given
by
\begin{equation}
 H_N = \frac{J}{N^{1-\alpha}}\sum_{i>j=1}^N \frac{1-S_iS_j}{|i-j|^{\alpha}}\quad,
\label{Hlattice}
\end{equation}
where $J>0$ and spins $S_i=\pm 1$ sit on a one-dimensional lattice
with unitary spacing. The $N^{\alpha-1}$ prefactor is introduced
in order to have an extensive energy. This model has been first
introduced by Dyson~\cite{dyson} and studied for the
``integrable'' case, $\alpha>1$, in the canonical ensemble without
the $N^{\alpha-1}$ prefactor. We will show that it is possible to
obtain an exact microcanonical solution using large deviation
theory, when $0\leq \alpha<1$. This solution can be easily
generalized to lattices of larger dimension. In a preliminary
paper~\cite{junext01}, a microcanonical solution of this model was
presented without rigorously proving  the exactness of the
mean-field limit.

The study of this model will also give us the opportunity to
emphasize the important role played by boundary conditions when
the interactions are long-range.  We will indeed consider both
free and periodic boundary conditions. In the latter case, $|i-j|$
is the minimal distance along the circle where one identifies the
first and the $(N+1)$-th lattice point.

In the solution we will adopt the same scheme described in
Section~\ref{Methode_generale}.

\textbf{Step 1:} The Hamiltonian $H_N$ cannot be rewritten exactly
using a finite dimensional global variable. We overcome this
difficulty by defining a coarse-grained function.  Let us divide
the lattice in $K$ boxes, each with $n=N/K$ sites, and let us
introduce the average magnetization in each box $m_k$, $k=1\dots
K$. In the limit $N\to \infty$, $K\to \infty$, $K/N \to 0$, the
magnetization becomes a continuous function $m(x)$, of the $[0,1]$
interval. After a long but straightforward calculation, described
in Appendix C, we can show that it is possible to express $H_N$ as
a functional of $m(x)$:
\begin{eqnarray}
H_N =  N H[m(x)] +o(N)  \quad, \label{approxH1}
\end{eqnarray}
where
\begin{eqnarray}
H[m(x)] = \frac{J}{2} \int_0^1\! dx \int_0^1 \!dy \frac{1-m(x)
m(y)}{|x-y|^{\alpha}}\quad.\label{formulepourh}
\end{eqnarray}
The estimation is uniform on all configurations.
\bigskip

\textbf{Step 2:} The probability to get a given magnetization
$m_k$ in the $k$-th box from all {\it a priori} equiprobable
microscopic configurations obeys a local large deviation principle
$P(m_k)\propto \exp[{n s(m_k)}]$, with
\begin{equation}
s(m_k)=-\frac{1+m_k}{2}\ln{\frac{1+m_k}{2}}-
\frac{1-m_k}{2}\ln{\frac{1-m_k}{2}}\quad.
\label{formuleentropie}
\end{equation}
Since the microscopic random variables in the different boxes are
independent and no global constraints has yet been imposed, the
probability of the full global variable  $(m_1,\ldots,m_K)$ can be
expressed in a factorized form as
\begin{eqnarray}
P(m_1,m_2,\ldots,m_K) & = & \prod_{i=1}^KP(m_i)
 \simeq  \prod_{i=1}^Ke^{\displaystyle n s(m_i)}=\exp\left[{ nK\sum_{i=1}^K
 \frac{s(m_i)}{K}}\right]
 \simeq e^{\displaystyle N S[m(x)]} \quad, \label{largedev2}
\end{eqnarray}
where $S[m(x)]=\int_0^1 s(m(x))\: dx$ is the entropy functional
associated to the global variable $m(x)$. Large deviation
techniques rigorously justify these
calculations~\cite{refnewellis}, proving that  entropy is
proportional to~$N$, also in the presence of long-range
interactions. This result is independent of the specific model
considered; it applies, for instance, also to the long-range XY
spin model studied by Salazar \emph{et al.}~\cite{Salazar02}.

\bigskip

 \textbf{Step 3:} It is now straightforward to formulate the
variational problem in the microcanonical ensemble
\begin{equation}
\label{isingmicro} S(\varepsilon)=\sup_{m(x)}\left( S[m(x)]\:
\left| \varepsilon=H[m(x)]\right.
\right)\quad.
\end{equation}

\begin{figure}[htbp]
\resizebox{0.5\textwidth}{!}{\includegraphics{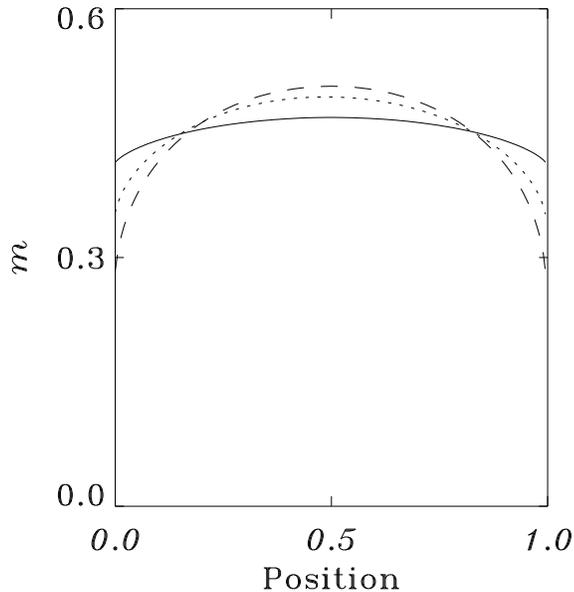}}
\caption{ Equilibrium magnetization profile  for the
$\alpha$-Ising model
  with free boundary conditions at an energy density
  $\varepsilon=0.1$ for $\alpha=0.2$ (solid line), $\alpha=0.5$ (dotted line)
  and $\alpha=0.8$ (dashed line).  }\label{magnprofile_fig}
\end{figure}

Let us remark that the optimization problem~(\ref{isingmicro}) has
to be solved in a functional space. In general, this has to be
done numerically, taking into account boundary conditions. In this
paper, we consider only free and periodic boundary conditions. In
the former case, the only available solutions are numerical. An
example of a maximal entropy magnetization profile obtained for
free boundary conditions is shown in Fig.~\ref{magnprofile_fig}
for different values of $\alpha$. In the following of this
section, we will treat the periodic boundary conditions case, for
which analytical result can be obtained.

\begin{figure}
\resizebox{0.95\textwidth}{!}{\includegraphics{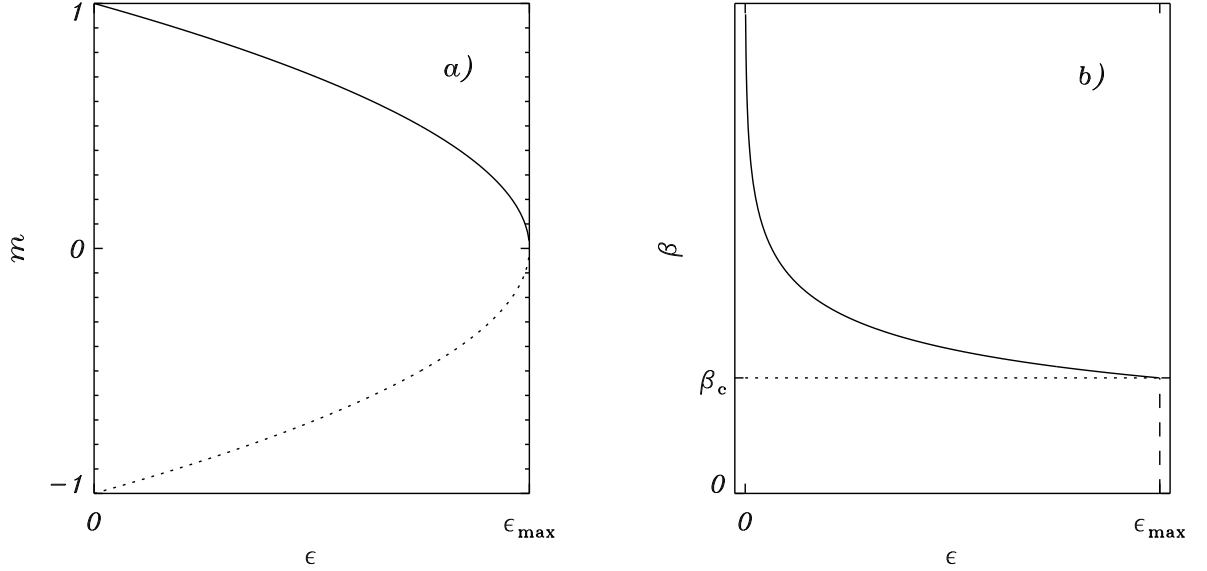}}
\caption{a) Equilibrium magnetization in the allowed energy range
in the microcanonical ensemble for the $\alpha$-Ising model with
$\alpha=0.5$; the negative branch is also reported with a dotted
line. b) Inverse temperature versus energy in the microcanonical
ensemble (solid line). The canonical ensemble result superposes to
the microcanonical one in the interval $[\beta_c,\infty]$ and is
represented by a dashed line for $\beta\in[0,\beta_c]$. $\beta_c$
is then the inverse critical temperature in the canonical
ensemble. In the microcanonical ensemble, no phase transition is
present.} \label{mbetaising}
\end{figure}

In the periodic boundary case, the distance $|x-y|$ in the energy
(\ref{Hlattice}) is defined as the minimal one on the circle,
obtained when the two boundaries of the interval $[0,1]$ are
identified. Both entropy and free energy can be obtained in
analytical form for homogeneous magnetization profiles.

In Appendix~D, we prove that for
$\beta<\beta_c=(1-\alpha)/(J2^\alpha)$ there is a unique global
maximum of~$S$, corresponding to a constant zero magnetization
profile. The variational problem (\ref{isingmicro}), where $S$ is
defined in Eq.~(\ref{formuleentropie}), leads to the
 equation
\begin{equation}
\label{isingmicro3} \tanh^{-1}\left(m(x)\right)={\beta\,
J}\int_0^1\frac{m(y)}{|x-y|^{\alpha}} \:dy\quad,
\end{equation}
where $\beta$ is a Lagrange multiplier. For $\beta>\beta_c$, we
restrict ourselves to solutions with constant magnetization
profiles, {\it i.e.}~$m(x)=m$. We  prove in Appendix D that these
solutions are locally stable, i.e. close non constant profiles
have a smaller entropy. Moreover, these are the only solutions
when $\alpha=0$, since the right-hand-side of Eq.
(\ref{isingmicro3}) is then independent of $x$. For constant
profiles, the relation between energy and magnetization is
\begin{equation}
\label{isingeneergy3} \varepsilon=\varepsilon_{\text{max}}
\left(1-m^2\right)\quad,
\end{equation}
where we have used $\int_0^1dx
|x-y|^{-\alpha}=2^\alpha/(1-\alpha)$ and
$\varepsilon_{\text{max}}=1/(2\beta_c)$. Hence, fixing the energy
implies fixing the magnetization and, consequently, the Lagrange
multiplier $\beta$ in Eq.~(\ref{isingmicro3}). Expressing the
magnetization in terms of the energy in the entropy formula
(\ref{formuleentropie}) allows to derive the caloric curve
${\beta}= d S/d \varepsilon$. The consistency equation
(\ref{isingmicro3}) has always a non vanishing magnetization
solution in the whole energy range $[0,\varepsilon_{\text{max}}]$:
this is reported in Fig.~\ref{mbetaising}(a). The caloric curve is
shown in Fig.~\ref{mbetaising}(b) with full line. The limit
temperature $\beta_c$ is attained at zero magnetization, which is
a boundary point.

In the canonical ensemble, one has to solve the variational
problem (\ref{varprobleme_cano}). This leads to exactly the same
consistency equation (\ref{isingmicro3}), where the Lagrange
multiplier is replaced by the inverse temperature $\beta$. Solving
this consistency equation on the full positive $\beta$ axis, one
finds a zero magnetization for $\beta<\beta_c$ and non vanishing
one for $\beta>\beta_c$. The caloric curve is obtained taking the
derivative $\varepsilon=d F/ d \beta$. The graph of this function
is reported in Fig. \ref{mbetaising}(b) and superposes to the
microcanonical caloric curve from infinity down to $\beta_c$ while
it is represented by the dashed line for $\beta<\beta_c$.

It follows that in the region $[0,\beta_c]$, the two ensembles are not
equivalent. In this case, a single microcanonical state at
$\varepsilon_{\text{max}}$ corresponds to many canonical states
with canonical inverse temperatures in the range $[0,\beta_c[$.
Thus, in such a case, the canonical inverse temperature is not
equal to the microcanonical one. In the microcanonical ensemble,
the full high temperature region is absent and, therefore, no
phase transition is present or, in other terms, the phase
transition is at the boundary of the accessible energy values. The
entropy is always concave, hence no inequivalence can be present
in the allowed energy range, apart from the boundaries. This
situation is called partial equivalence~\cite{Ellis99,Kastner}.
This ensemble inequivalence persists for all $\alpha$ values below
one, and is removed only for $\alpha=1$ when
$\varepsilon_{\text{max}}\to\infty$ and $\beta_c\to0$: the phase
transition is not present in both ensembles and the system is
always in its magnetized phase.

The main drawback of this analysis is the difficulty to obtain
analytical solutions of Eq.~(\ref{isingmicro3}) for non constant
magnetization profiles, which is the typical situation when
boundary conditions are not periodic. However, we have shown that,
for periodic boundary conditions, constant magnetization profiles
are locally stable (see Appendix~D), but the proof of non existence of 
generic magnetization profiles with larger entropy, and hence the 
global stability analysis, remains to be done.

An advantage of the method we have exposed is its flexibility and
applicability to more complex models. For instance, some results
of the kind presented here have been already obtained for the
$\alpha$-Blume-Emery-Griffiths
model~\cite{barrethesis,MukamelHouches}.

\begin{figure}[htbp]
\resizebox{0.48\textwidth}{!}{\includegraphics{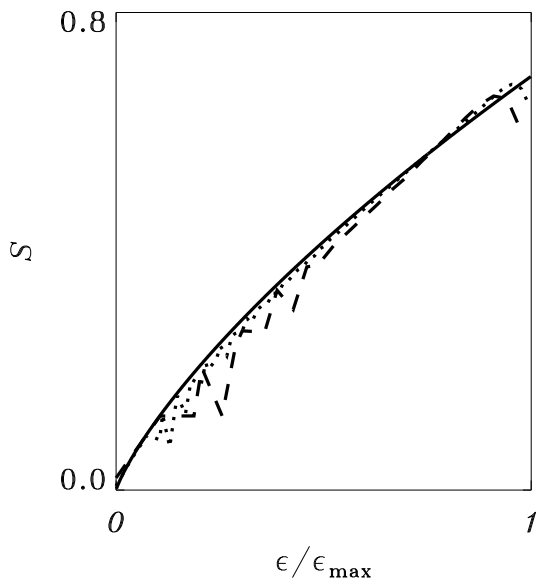}}
\caption{Entropy density versus energy density for the
$\alpha$-Ising model but with periodic boundary conditions and
with $\alpha=0.8$.  The solid curve represents the theoretical
solution, whereas the dashed and dotted lines are determined by
microcanonical Monte-Carlo simulations with $N=34$ and $N=100$,
respectively (data provided by R. Salazar).
}\label{entroprofile_fig}
\end{figure}

It is also interesting to check how fast one reaches the $N\to
\infty$ solution as one increases the number of spins $N$.
Fig.~\ref{entroprofile_fig} shows entropy density as a
function of energy density for the $\alpha$-Ising model with
$\alpha=0.8$ and periodic boundary conditions. The figure
emphasizes that the asymptotic result is already accurate enough
for $N =100$.

\section{Conclusions and perspectives}
\label{remarques}

In this paper we have discussed examples of the application of
large deviation techniques to the study of the statistical
mechanics of infinite-range models, at equilibrium, in the
microcanonical and canonical ensembles. Besides that, we have
shown how to construct a mean-field Hamiltonian for the Ising
model in one dimension with $1/r^\alpha$ interaction
($0\leq\alpha<1$). The solution of simple toy models already shows
interesting ensemble inequivalence features, like negative
specific heat. Among these simple models, one should point out
those with continuous state variables, because the Hamiltonian
dynamics becomes accessible and one can then study also
non-equilibrium features. Remarkable is also the solution of the
Colson-Bonifacio model, which is believed to capture the
phenomenology of the saturated state.

It is important to emphasize that the method proposed in this
paper does not apply to all long-range interacting systems. In
particular, those for which statistical mechanics cannot be
reduced to a mean-field variational problem are excluded. As
presented in Section \ref{Methode_generale}, the method is
strongly dependent on the possibility to introduce global or
coarse-grained variables: examples are the averaged magnetization,
the total kinetic energy, {\em etc}. The coarse-grained variables
allow one to describe structures whose size is of the order of the
total size of the system; however they may be insufficient to
characterize the effect of short-range interaction.  A typical
example is the Ising model with attractive short-range
interactions and repulsive long-range couplings studied by
Grousson {\em et al.} \cite{Tarjus01}. The creation of large scale
structures needs an infinite energy, because of the repulsive
long-range component. However, the competition between repulsion
and attraction can lead to interesting structures, such as
alternating positive and negative magnetized stripes. Indeed,
despite the long-range character of the interaction, this system
is additive and these structures are therefore compatible with the
predicted zero magnetization at large scale. Still in the context
of canonical ensemble, worth quoting are the results of
Kardar~\cite{Khardar} who, analyzing an Ising model with both
short-range and long-range interactions was able to derive the
exact free energy by a minimization procedure. Obtaining similar
results for the microcanonical entropy would be extremely
important.

The application of the techniques described in this paper to more
realistic $N$-body systems is an important issue. The extension to
wave-particle  interactions~\cite{Escande} should not be too
difficult. The mean-field description of the two-dimensional point
vortices model~\cite{Onsager49} has been rigorously obtained in a
series of papers~\cite{Kiessling97,kiesling,cagliotti}. On the
contrary, due to the strong short-distance singularity, similar
results for self-gravitating systems~\cite{Lynden68,Padmanabhan90}
have not been obtained and are a challenging current issue. What
is usually done is: i) in equilibrium, to conjecture the validity
of the mean-field description; ii) out-of-equilibrium, to consider
the Vlasov-Poisson equation
 as a good approximation of
the short time dynamics. Once the mean-field description is
introduced at finite time (Euler equation, Vlasov-Poisson
equation), the statistical mechanics can be derived using large
deviation techniques~\cite{Michel94,Ellis99}. However, it is not
equivalent to the original statistical mechanics of the $N$-body
system. Examples of this inequivalence, due to the exchange of the
two limits $t\to \infty$ and $N\to \infty$, are given in
Refs.~[\onlinecite{yoshi,FELPRE}] for the HMF and the
Colson-Bonifacio model.

\bigskip

{\bf Acknowledgement:} We would like to thank J. Michel for very
helpful advices, R.~Salazar for providing us the numerical data
used in Fig.~\ref{entroprofile_fig} and A. Alastuey, A. Antoniazzi,
P. Cipriani and H. Touchette for
discussions. This work has been partially supported by the French
Minist{\`e}re de la Recherche grant ACI jeune chercheur-2001
N$^\circ$ 21-31 and the R{\'e}gion Rh{\^o}ne-Alpes for the
fellowship N$^\circ$ 01-009261-01. It is also part of the
contract COFIN03 of the Italian MIUR {\it Order and chaos in
nonlinear extended systems}.

\section*{Appendix A: Alternative derivation of the microcanonical
solution of the HMF model} \label{appendix} The microcanonical
solution of the HMF model~(\ref{Ham_HMF}) can be alternatively
obtained using the traditional method applied for self-gravitating
systems \cite{binneytremaine}.  Denoting by $K$ and $V$
kinetic and potential energy, respectively, the number of
microscopic configurations corresponding to the energy $E$ is
given by
\begin{eqnarray}
  \Omega_N(E)&=&\int\prod_i  dp_i d\theta_i\, \delta(E-H_N) \\
&=&\int\prod_i  dp_i d\theta_i\underbrace{\int dK  \,
\delta\left(K-\sum_i\frac{p_i^2}{2}\right)}_{=1}\,
  \delta\left(E-K-V(\{\theta_i\})\right) \\
&=&\int dK \underbrace{\int\prod_i  dp_i   \,
  \delta\left(K-\sum_i\frac{p_i^2}{2}\right)}_{ \Omega_{\rm kin}(K)}\,
 \underbrace{ \int\prod_i  d\theta_i \delta
\left(E-K-V(\{\theta_i\})\right)}_{ \Omega_{\rm conf}(E-K)}\quad.
\label{eq:tradwaysuite}
\end{eqnarray}
The factor $\Omega_{\rm kin}$ is classical and corresponds to the
volume of the hypersphere with radius $R=\sqrt{2K}$ in $N$
dimensions: its expression is  $\Omega_{\rm
kin}={2\pi^{N/2}}/{\Gamma(1+N/2)}$.  Using the asymptotic
expression of the $\Gamma$-function, $\ln \Gamma(N)\simeq N\ln
N-N$, one obtains
\begin{eqnarray} \Omega_{\rm kin}\left(K\right)&\stackrel{N\to+\infty}{\sim}&
\exp\left(\frac{N}{2}\left[1+\ln \pi-\ln \frac{N}{2}+\ln
(2K)\right]  \right)\\
&=&  \exp\left(\frac{N}{2}\left[1+\ln (2\pi)+\ln u\right]
\right)\quad, \label{omegazero}
\end{eqnarray}
where $u=2K/N$. Defining the configurational entropy per particle
$s_{\rm
  conf}(\widetilde V)=(\ln \Omega_{\rm conf}(N\widetilde V))/N$,
where $\widetilde V=(E-K)/N$, Eq.~(\ref{eq:tradwaysuite}) can be
rewritten as
\begin{eqnarray}
\Omega_N(\varepsilon N)&\stackrel{N\to+\infty}{\sim}& \frac{N}{2}
\int du \exp\left[{N}\left(\frac{1}{2}+\frac{\ln (2
\pi)}{2}+\frac{1}{2}\ln u +s_{\rm conf}(\widetilde V)\right)
\right]\quad. \label{eq:tradwaysuitetekkr}
\end{eqnarray}
Hence, solving the integral in the saddle point approximation,
gives the entropy modulo a trivial constant
\begin{eqnarray}
S(\varepsilon)&=&\lim_{N\to +\infty}\frac{1}{N}\ln \Omega_N(\varepsilon N)\\
&=&\frac{1}{2}\sup_u\left[\frac{1}{2}\ln u +s_{\rm
conf}(\widetilde V) \right] \label{final}
\end{eqnarray}
to which Eq. (\ref{casmicrocanonique}) reduces once the $\sup$ on
$M$ is performed. Indeed, this expression of the entropy assumes
the knowledge of the configurational entropy $s_{\rm conf}$, which
is determined by solving the  extremal condition in Eq.
(\ref{sconf}), and is restricted to total vanishing momentum
$v=0$. Hence, the method given in the text is slightly more
general.

\section*{Appendix B: Mapping the FEL model onto HMF} \label{appendixFEL}

The microcanonical solution of the FEL model~(See Eq.
(\ref{eq:Hamiltonien})) can be expressed in terms of the HMF
Hamiltonian using the Laplace representation of the Dirac
$\delta$-function and a Gaussian integration. After performing the
change of variables
 $\widetilde{\theta}_i=\theta_i-\varphi$, the microcanonical volume of the FEL
is given by
\begin{eqnarray}
\Omega(E)&=&\int\!\!\!\int\!\!\!\int\!\!\prod_i  dp_i
d\widetilde{\theta}_i dA\, \delta(E-H_N)
\label{eq:tradwaysuiteFEL}\\
&=&\int\!\!\!\int\!\!\!\int\!\!\prod_i  dp_i d\widetilde{\theta}_i
dA\,\frac{1}{2i\pi}\int_\Gamma d\lambda\,
e^{\displaystyle\lambda(E-H_N)}\quad,
\end{eqnarray}
where $\Gamma$ is a path on the complex $\lambda$-plane, going
from $-i\infty$ to $+i\infty$, which crosses the real axis at a
positive value.

Introducing the FEL Hamiltonian
\begin{equation}
\Omega(E)=\int\prod_i  dp_i \,
e^{\displaystyle\lambda\left(E-\sum_{j=1}^N\frac{p_j^2}{2}\right)}\int\!\!\!
\int\prod_i d\widetilde{\theta}_i dA\,\frac{1}{2i\pi}\int_\Gamma
d\lambda\, e^{\displaystyle\lambda(N \delta A^2
-2NA\widetilde{M}_y)}\quad,
\end{equation}
and performing the Gaussian integral over the field variable $A$,
one gets
\begin{eqnarray}\Omega(E)&=&\int\prod_i  dp_i \int\prod_i d\widetilde{\theta}_i
\frac{1}{2i\pi}\int_\Gamma d\lambda\
e^{\displaystyle\lambda\left(E-\sum_{j=1}^N\frac{p_j^2}{2} - N
\frac{\widetilde{M}_y^2}{\delta}-\frac{1}{2}\ln\left[ \lambda N
(-\delta)\right] \right)} \frac{\sqrt{\pi}}{2}\\
&=&\int\prod_i dp_i \int\prod_i d\widetilde{\theta}_i
\delta\left(E-\sum_{j=1}^N\frac{p_j^2}{2} - N
\frac{\widetilde{M}_y^2}{\delta}-\frac{1}{2}\ln\left[ \lambda N
(-\delta)\right] \right) \frac{\sqrt{\pi}}{2}\quad.
 \label{eq:tradwaysuiteFELbb}
\end{eqnarray}
In the large $N$-limit, one can neglect all constants and the $\ln
N$ term in the argument of the Dirac $\delta$, obtaining the
microcanonical volume  for the Hamiltonian
\begin{equation}
H_N=\sum_{j=1}^N\frac{p_j^2}{2}
+N\frac{\widetilde{M}_y^2}{\delta}\quad.
\end{equation}
Hence, for negative values of the parameter $\delta$, solving the
microcanonical problem for the FEL
Hamiltonian~(\ref{eq:Hamiltonien}) is formally equivalent to
obtaining the solution of the HMF Hamiltonian~(\ref{Ham_HMF2})
with $C=2/\delta$ and $M_x=0$.

\section*{Appendix C: Derivation of formula (\ref{approxH1}) for the
  $\alpha$-Ising model} \label{appendixBtheseJulien}

Denoting by $\Sigma_N=\{-1,1\}^N$ the phase space of the
$\alpha$-Ising model, its Hamiltonian is defined as a function of
$\Sigma_N $ to $\mathbb{R}$
\begin{eqnarray}
H_N & :&  \Sigma_N\to \mathbb{R} \\
    &  &  \omega_N \: \: \:\mapsto J \sum _{i> j} \frac{1-S_iS_j}{|i-j|^
{\alpha}}~,
\end{eqnarray}
where $\omega_N=\left(S_1,\ldots, S_N\right)$ is a given
microscopic configuration.

The coarse-graining operator  $Y_{N,K}$ divides the lattice into
$K$ boxes of size $n=N/K$ and defines a locally averaged
magnetization  $m_k$ in the  $k^{\text{th}}$ box $B_k$. The
coarse-grained magnetization is then a step function, which takes
a constant value
\begin{equation}
m_k=\frac{1}{n}\sum_{i\in B_k} S_i\quad,
\end{equation}
in $B_k$. It is convenient to introduce the operator
\begin{eqnarray}
Y_{N,K} & : &  \Sigma_N \to {L^2}([0,1])\quad, 
\end{eqnarray}
which maps  the configuration space to the coarse-grained
magnetization. The length of the lattice is then renormalized to
the interval $[0,1]$ and, in the limit $N\to\infty$, this operator
defines a continuous magnetization profile~$m(x)$. The limit of
the number of boxes $K\to\infty$ is taken in such a way that the
number of sites per box diverges $N/K\to\infty$, as already
mentioned in the text. In the following, we will use free boundary
conditions (The choice of periodic boundary conditions would
change only some details of the calculation).

The energy as a function of the coarse-grained magnetization is
defined as
\begin{equation}
\widetilde{h}[Y_{N,K}(\omega_N)]=\sum_{k,l=1}^K
\frac{J}{2}(1-m_km_l)\,d_{kl}\quad,
\end{equation}
where
\begin{equation}
d_{kl}=\int_{{(k-1)}/{K}}^{\,{k}/{K}}
\int_{{(l-1)}/{K}}^{\,{l}/{K}} dx\:dy
\frac{1}{|x-y|^{\alpha}}\quad.
\end{equation}
Hence, the composition of the operator $Y_{N,K}$ with the
functional $\widetilde{h}$ allows to define a map, which
associates to each microscopic configuration $\omega_N\in\Sigma_N$
a given energy. Our aim is to find a series $K(N)$ such that
\begin{equation}
\label{convuniforme} \lim_{N\to\infty} \sup_{\omega_N\in\Sigma_N}
\left|\frac{H_N(\omega_N)}{N^{2-\alpha}}-\widetilde{h}[Y_{N,K(N)}(\omega_N)]
\right|=0\quad.
\end{equation}
This is what we mean by uniform convergence of the Hamiltonian
$H_N(\omega_N)$ to its functional form ${H}[m(x)]$, defined in
formula~(\ref{formulepourh}).

The proof relies on the long-range character of the interaction
($0\leq\alpha <1$). It is straightforward but lengthy.

Let us
first express $\widetilde{h}[Y_{N,K(N)}(\omega_N)]$ directly as a
function of the spin variables $S_i$
\begin{equation}
\widetilde{h}[Y_{N,K(N)}(\omega)]=\sum_{k,l=1}^K
\frac{d_{kl}}{n^2} \sum_{\substack{i\in B_k\\j\in B_l}}
\frac{J}{2}(1-S_iS_j)\quad. \label{approxcontinue}
\end{equation}

In order to compare expression~(\ref{approxcontinue}) with
$H_N/N^{2-\alpha}$, let us first introduce the reduced Hamiltonian
\begin{equation}
g_{N,K}=\frac{1}{N^{2-\alpha}}\sum_{\substack{k,l=1\\ l\neq k}}^K
\sum_{\substack{i\in B_k\\j\in B_l}}
\frac{J}{2}\frac{(1-S_iS_j)}{|nk-nl|^{\alpha}}\quad,
\end{equation}
which is in fact the original Hamiltonian $H_N/N^{2-\alpha}$ where
the distance between two sites is approximated by the distance
among the boxes to which they belong.  Using the triangular
inequality, one gets
\begin{equation}
\left|\frac{H_N(\omega_N)}{N^{2-\alpha}}
-\widetilde{h}[Y_{N,K(N)}(\omega_N)]\right|\leq
\left|\frac{H_N(\omega_N)}{N^{2-\alpha}} -g_{N,K}\right|+
\left|g_{N,K}-\widetilde{h}[Y_{N,K(N)}(\omega_N)]\right|\equiv
A+B\quad.
\end{equation}
We will show how both $A$ and $B$ can be bounded from above by
quantities which vanish when $N\to \infty$, under the hypothesis
that $0\leq \alpha<1$.

\subsubsection{Upper bound of $A$} Let us first rewrite
$H_N/N^{2-\alpha}$ as a sum of two terms: the first contains
contributions from sites which belong to different boxes, the
second from those which are in the same box,
\begin{equation}
\frac{H_N}{N^{2-\alpha}} = \frac{1}{N^{2-\alpha}}
\sum_{\substack{k,l=1\\ l\neq k}}^K \sum_{\substack{i\in B_k\\
j\in B_l}} \frac{J}{2}\frac{(1-S_iS_j)}{|i-j|^{\alpha}}
+\frac{1}{N^{2-\alpha}}\sum_{k=1}^K\sum_{\substack{i,j \in B_k}}
\frac{J}{2}\frac{(1-S_iS_j)}{|i-j|^{\alpha}}\quad.
\end{equation}
Then
\begin{eqnarray}
A &\leq& \frac{1}{N^{2-\alpha}}\left|\sum_{\substack{k,l=1\\ l\neq
k}}^K
\sum_{\substack{i\in B_k\\
j\in
B_l}}\frac{J}{2}(1-S_iS_j)\left(\frac{1}{|i-j|^{\alpha}}-\frac{1}
{|nk-nl|^{\alpha}}\right)\right| +\left|\frac{1}{N^{2-\alpha}}
\sum_{k=1}^K\sum_{\substack{i,j \in B_k}}
\frac{J}{2}\frac{(1-S_iS_j)}{|i-j|^{\alpha}}\right|\equiv A_1+A_2
\quad. \label{eq:majoration}
\end{eqnarray}
The bound of  $A_2$ is easy to find since  $J(1-S_iS_j)/2\leq J$
and $|i-j|^{\alpha}\geq 1$ for $\alpha\in[0,1]$. One obtains
\begin{equation}\label{formA2}
  A_2\leq J\frac{Kn^2}{N^{2-\alpha}}=J\frac{N^{\alpha}}{K}\quad.
\end{equation}
This quantity vanishes in the large $N$ limit if $K$ diverges
faster than $N^{\alpha}$ (and certainly slower than $N$). This is
the first point where the long-range nature of the interaction is
used.

In order to bound $A_1$ let us divide the first sum in
formula~(\ref{eq:majoration}) into three parts : $k-l>1$, $k-l<-1$
and $|k-l|=1$. The last part, $|k-l|=1$, can be bounded from above
by $2JN^{\alpha}/K$, which vanishes in the large $N$ limit. The
two remaining parts are symmetric, hence we treat only the case
$k-l>1$. Since $i\in B_k$ and $j\in B_l$, this implies that
 $n(k-l-1)<i-j< n(k-l+1)$. Hence,
\begin{equation}
\frac{1}{n^{\alpha}}
\frac{1}{(k-l+1)^{\alpha}}<\frac{1}{|i-j|^{\alpha}}
<\frac{1}{n^{\alpha}}\frac{1}{(k-l-1)^{\alpha}}\quad.
\end{equation}
Substracting $1/|nk-nl|^{\alpha}$, one gets
\begin{equation}
\frac{1}{n^{\alpha}}\left(\frac{1}{(k-l+1)^{\alpha}}
-\frac{1}{(k-l)^{\alpha}}\right)
<\frac{1}{|i-j|^{\alpha}}-\frac{1}{n^{\alpha}(k-l)^{\alpha}}
<\frac{1}{n^{\alpha}}\left(\frac{1}{(k-l-1)^{\alpha}}
-\frac{1}{(k-l)^{\alpha}}\right)\quad.\label{substracting}
\end{equation}
This leads to
\begin{equation}
\left|\frac{1}{|i-j|^{\alpha}}-\frac{1}{n^{\alpha}(k-l)^{\alpha}}\right|
<\frac{1}{n^{\alpha}}\left(\frac{1}{(k-l-1)^{\alpha}}
-\frac{1}{(k-l)^{\alpha}}\right)\quad.
\end{equation}
Thus
\begin{eqnarray}
\label{sommekl}
\frac{1}{N^{2-\alpha}}\sum_{\substack{k,l=1\\k-l>1}}^K\sum_{\substack{i\in
B_k\\j\in B_l}}
\left|\frac{1}{|i-j|^{\alpha}}-\frac{1}{n^{\alpha}(k-l)^{\alpha}}\right|
&<&
\frac{1}{N^{2-\alpha}}\frac{n^2}{n^{\alpha}}\sum_{\substack{k,l=1\\k-l>1}}^K
\left[\frac{1}{(k-l-1)^{\alpha}}-\frac{1}{(k-l)^{\alpha}}\right]\\
&=&\frac{n^{2-\alpha}}{N^{2-\alpha}}\sum_{k=3}^{K}\sum_{l=1}^{k-2}
\left[\frac{1}{(k-l-1)^{\alpha}}-\frac{1}{(k-l)^{\alpha}}\right]\\
&=&\frac{1}{K^{2-\alpha}}\sum_{k=3}^{K}\left[1-\frac{1}{(k-1)^{\alpha}}
\right] <\frac{K}{K^{2-\alpha}}\quad.
\end{eqnarray}
Hence, this part of $A_1$ is bounded by $JK^{\alpha-1}$, which
vanishes in the large $N$ limit. The symmetric part of $A_1$
$k-l<-1$ can be treated exactly in the same way.

\subsubsection{Upper bound of  $B$} Dividing the sum in two parts,
one for sites in different boxes and the other for sites in the
same box, and using the triangular inequality, one gets
\begin{eqnarray}
B &\leq&\frac{1}{N^{2-\alpha}} \left|\sum_{\substack{k,l=1\\ l\neq
k}}^N\sum_{\substack{i\in B_k\\j\in B_l}}
\frac{J}{2}(1-S_iS_j)\left(\frac{1}{|nk-nl|^{\alpha}}-\frac{d_{kl}}{n^2}\right)\right|
 +\left|\sum_{k}\frac{d_{kk}}{n^2}\sum_{i,j\in B_k}\frac{J}{2}(1-S_iS_j)
\right|\equiv B_1+B_2\quad.
\end{eqnarray}
In order to estimate the size of $B_2$, one has first to evaluate
$\sum_k d_{kk}$. Indeed,
\begin{eqnarray}
d_{kl} 
& = & \int_{0}^{\,{1}/{K}} \! \int_{0}^{\,{1}/{K}}
dx\:dy \frac{1}{|x-y+\frac{k-l}{K}|^{\alpha}}
= \frac{1}{K^{2-\alpha}}\int_0^1 \! \int_0^1 du\:dv
\frac{1}{|u-v+k-l|^{\alpha}}\quad.\label{formuledkl}
\end{eqnarray}
Therefore  $\sum_k d_{kk}\sim K^{\alpha-1}$. Hence, $B_2\leq
JK^{\alpha-1}$, which vanishes in the large $N$-limit.

For what $B_1$ is concerned, exchanging the modulus with the sums
and using the expression for
$d_{kl}$~(formula~(\ref{formuledkl})), one obtains
\begin{eqnarray}
B_1 &\leq& J\sum_{\substack{k,l=1\\l\neq k}}^N
\left|\frac{n^2}{N^{2-\alpha}n^{\alpha}|k-l|^{\alpha}}-d_{kl}\right|
 \leq J\sum_{\substack{k,l=1\\l\neq k}}^N\frac{1}{K^{2-\alpha}}
\left|\int_0^1 \int_0^1 du\:dv \frac{1}{|u-v+k-l|^{\alpha}}
-\frac{1}{|k-l|^{\alpha}}\right|\quad.
\end{eqnarray}
Similarly as for $A_1$, one divides the sum in three parts
$k-l>1$, $k-l<-1$ and $|k-l|=1$. The last part gives a term of
order   $JK^{\alpha-1}$, which vanishes in the large $N$ limit. If
$k-l>1$, analogously to Eq.~(\ref{substracting}), one obtains
\begin{equation}
\begin{split}
\frac{1}{(k-l+1)^{\alpha}}-\frac{1}{(k-l)^{\alpha}} < \int_0^1 \!
\int_0^1 du\:dv \frac{1}{|u-v+k-l|^{\alpha}}-\frac{1}
{|k-l|^{\alpha}}
 < \frac{1}{(k-l-1)^{\alpha}}-\frac{1}{(k-l)^{\alpha}}~\quad,
\end{split}
\end{equation}which leads to
\begin{equation}
\left|\int_0^1 \! \int_0^1 du\:dv \frac{1}{|u-v+k-l|^{\alpha}}-
\frac{1}{|k-l|^{\alpha}}\right|
<\frac{1}{(k-l-1)^{\alpha}}-\frac{1}{(k-l)^{\alpha}}\quad.
\end{equation}
Summing this term over $k$ and $l$ gives again a factor of order
$K$. Then, this part is also bounded from above by a factor
$JK^{\alpha-1}$. The symmetric part $k-l<1$ is equally treated.

All terms in $B$ are bounded by $JK^{\alpha-1}$, and therefore
vanish in the large $N$ limit. This concludes the proof of
formula~(\ref{convuniforme}) and assures that $\widetilde{h}$
converges to $H$.

We would like to emphasize that throughout of this derivation, the
only place where we use the Ising interaction of the model is
where we bound $J(1-S_iS_j)/2$ by $J$. It is then easy to realize
that the method can be generalized to other models like Potts,
Blume-Emery-Griffiths, XY, with spatially decaying interactions.
On the contrary, the presence of the lattice is crucial, because
it avoids microscopic configurations where the state variables
concentrate on single point, leading to a divergence.

\section*{Appendix D: Proof of stability of constant magnetization profiles for the
$\alpha$-Ising model with periodic boundary conditions}
\label{appendixIsing}

We prove here the stability of constant magnetization profiles for
the $\alpha$-Ising model in the context of the canonical ensemble.
This also implies stability in the microcanonical ensemble. Let us
consider the free energy $F[m(x)]$ as a functional of $m(x)$
\begin{equation}
F[m(x)] =-S[m(x)] + \beta H[m(x)]\quad,
\end{equation}
where the entropy functional $S[m(x)]$ is defined in formula
(\ref{largedev2}). We will limit ourselves to periodic
boundary conditions.  Let us study the second variation of $F$
\begin{equation}
\delta^2 F(m(x))=-\int_0^1dx\,s''[m(x)]\,\delta m(x)^2 -\beta
\frac{J}{2}\int_0^1 dx \int_0^1 dy \frac{\delta m(x)\delta
m(y)}{|x-y|^\alpha}\quad,\label{delatF}
\end{equation}
where $s''$ is the second derivative of $s$. Let us express the
variation $\delta m(x)$ in Fourier components
\begin{equation}
\delta m(x) = \sum_{k=-\infty}^{+\infty} \delta m_k e^{2i\pi
kx}\quad,\label{Fourierexp}
\end{equation}
and define
\begin{equation}
c_k^\alpha = \int_0^1 dx\ \frac{e^{2i\pi kx}}{|x-y|^\alpha} = 2
\int_0^{1/2} dx \frac{\cos(2 \pi k x)}{|x|^\alpha}\quad,
\end{equation}
where, since we consider periodic boundary conditions, $|\cdot|$
denotes the distance on the circle. Using the inequality
$s''(m(x))\leq s''(0)$ and  replacing the Fourier expansion
(\ref{Fourierexp}) in Eq. (\ref{delatF}), we obtain
\begin{equation}
\delta^2 F(m(x)) \geq \sum_{k=-\infty}^{+\infty}\left(-s''(0)
 -\beta \frac{J}{2} c_k^\alpha \right)\delta
m_k^2\quad.\label{fouriersum}
\end{equation}
Remarking that all coefficients $c_k^\alpha$ are positive reals
and that for all $k$, $c_k^\alpha \leq
c_0^\alpha=2^\alpha/(1-\alpha)$, one can bound from above all the
coefficients in the Fourier sum (\ref{fouriersum})  by $(-s''(0) -
\beta Jc_0^\alpha/2) $.

Thus for $\beta<-2s''(0)/(Jc_0^\alpha)=\beta_c$, one has $\delta^2
F(m(x))>0$. In such a case, the free energy is strictly convex,
and hence its minimum is unique. This proves the global stability
of the state $m(x)=0$.

For $\beta>\beta_c$, we do not study global stability. Let us
however show the local stability of the uniform magnetization
state $m(x)=m$. Using Eq.~(\ref{delatF}), one has
\begin{equation}
\delta^2 F(m(x)) \geq \sum_{k=-\infty}^{+\infty}\left(-s''(m)
 -\beta\frac{ J}{2} c_k^\alpha \right)\delta
m_k^2\quad.\label{fouriersumbig}
\end{equation}
Using again the property $c_k^\alpha \leq c_0^\alpha$, we note
that all the Fourier coefficients of Eq.~(\ref{fouriersumbig}) are
bounded from above by $-s''(m) -\beta J c_0^\alpha/2$.
 Observing that this latter expression is the
second variation of the free energy for uniform magnetization
profiles, i.e. $(1-m^2(\beta))^{-1}-\beta/\beta_c$, the analysis
of Section~\ref{Exemple_alphaIsing} implies that this quantity is
non negative both for $m \neq 0$ and $\beta > \beta_c$ and for $m
= 0$ and $\beta \leq \beta_c$. This proves that close to a
constant magnetization profile, there is no non uniform
magnetization profile which gives a smaller free energy functional
$F[m(x)]$.


\begin{thebibliography}{99}
\bibitem{limitperformed}
For short-range interactions, the limit is performed by letting
both the number of particles and volume go to infinity, with a
fixed ratio between the two quantities.
\bibitem{Alternatively}Alternatively, one could have left  energy non-extensive and
accordingly rescale temperature.
\end{thebibliography}
\end{document}